\documentclass[a4paper,fleqn,usenatbib]{mnras}

\usepackage[T1]{fontenc}
\usepackage{ae,aecompl}

\usepackage{graphicx}	
\usepackage{amsmath}	
\usepackage{amssymb}	

\title[CO X-ray absorption detectability]{On the detectability of CO molecules in the Interstellar Medium via X-ray spectroscopy}

\author[Joachimi et al.]{
Katerine Joachimi,$^{1}$\thanks{E-mail: katerine.joachimi@ciens.ucv.ve}
Efra\'in~Gatuzz,$^{1,2}$\thanks{E-mail: egatuzz@mpa-garching.mpg.de}
Javier~A.~Garc\'ia$^{3}$ 
and Timothy~R.~Kallman$^{4}$ 
\\
$^{1}$Escuela de F\'isica, Facultad de Ciencias, Universidad
Central de Venezuela, PO Box 20632, Caracas 1020A, Venezuela\\
$^{2}$Max-Planck-Institut f\"ur Astrophysik,
85741 Garching bei M\"unchen, Germany\\
$^{3}$Harvard-Smithsonian Center for Astrophysics, Cambridge, MA, 02138,  USA\\
$^{4}$NASA Goddard Space Flight Center, Greenbelt, MD 20771, USA
}

\date{Accepted XXX. Received YYY; in original form ZZZ}

\pubyear{2016}

\begin{document}
\label{firstpage}
\pagerange{\pageref{firstpage}--\pageref{lastpage}}
\maketitle

\begin{abstract}
We present a study of the detectability of CO molecules in the Galactic interstellar medium using high-resolution X-ray spectra obtained with the {\it XMM-Newton} Reflection Grating Spectrometer. We analyzed 10 bright low mass X-ray binaries (LMXBs) to study the CO contribution in their line-of-sights. A total of 25 observations were fitted with the {\tt ISMabs} X-ray absorption model which includes photoabsorption cross-sections for O~{\sc I}, O~{\sc II}, O~{\sc III} and CO. We performed a Monte-Carlo (MC) simulation analysis of the goodness of fit in order to estimate the significance of the CO detection. We determine that the statistical analysis prevents a significant detection of CO molecular X-ray absorption features, except for the lines-of-sight toward XTE~J1718-330 and 4U~1636-53. In the case of XTE~J1817-330, this is the first report of the presence of CO along its line-of-sight. Our results reinforce the conclusion that molecules have a minor contribution to the absorption features in the O K-edge spectral region. We estimate a CO column density lower limit to perform a significant detection with {\it XMM-Newton} of $N({\rm CO})> 6$ $\times 10^{16}$ cm$^{-2}$ for typical exposure times.
\end{abstract}




\section{Introduction}

The physical conditions in the interstellar medium (ISM) can be studied through
the technique of high-resolution X-ray spectroscopy. Using a bright
astrophysical X-ray source as a background lamp, one can analyze the absorption
features that are imprinted in the spectra by the ISM material located between
the observer and the source. Due to their high energy, X-ray photons interact
not only with the atomic phase but also with molecules and dust. In this way,
high-resolution X-ray spectra provide a powerful method to study basic
properties of the ISM such as composition, degree of ionization, column
densities, and elemental abundances.

The cold phase of the ISM is composed mostly of molecular gas, predominantly
molecular hydrogen. However, there are no transitions in the H$_2$ molecule
that can be excited at low temperatures, and thus other tracers need to be
employed. Carbon monoxide (CO) is the next most abundant molecule \citep{wil70}
in the ISM.  The CO molecule can give rise to characteristic X-ray
absorption features the oxygen K-edge region (21-24\AA).

Multiple studies have been performed reporting the presence of multiple phases
in the ISM using low mass X-ray binaries (LMXBs) as X-ray sources
\citep{jue04,jue06,pin10,pin13,lia13,luo14}.
Regarding the molecular contribution to the
spectra, \citet{pin10} searched for CO molecular absorption by
analyzing the {\it XMM-Newton} spectrum of the LMXB GS~1826-238, finding an
upper limit for the CO column density of $< 0.4\times10^{17}$cm$^{-2}$. In a
similar way, \citet{pin13} analyzed the local ISM in the lines-of-sight toward nine
LMXBs, reporting CO column densities of $0.02-0.7\times10^{17}$cm$^{-2}$.
However, an adequate modeling of the atomic component is imperative before
trying to model the contribution due to molecules or dust. In this sense,
\citet{gat13a,gat13b,gat14,gat15} and \citet{gat16a} have conducted a
sequential analysis of the galactic ISM using high-resolution X-ray spectra
obtained with {\it Chandra} and {\it XMM-Newton} observatories, finding a
satisfactory modeling of the O K-edge region using only atomic and ionic absorbers.

In this paper we present a study of the detectability of CO molecules in the
ISM by analyzing the {\it XMM-Newton} spectra of 10 bright LMXBS. In modeling
these spectra, we use the most-up-to date atomic data for atomic oxygen in
combination with experimental measurements of CO photoabsorption cross-section
\citep{bar79}. The outline of this paper is as follows.  In
Section~\ref{sec_dat} we describe the reduction of the observational data.  In
Section~\ref{sec_ofit} we describe the O K-edge model which is used to fit the
spectra, and provide details concerning the atomic and molecular database. In
Section~\ref{sec_dis} we discuss in detail the main results.
Section~\ref{sec_previous} is dedicated to compare our results with previous
works. Finally, in Section~\ref{sec_con} we summarize our main conclusions.
%

\begin{table*}
\caption{\label{tab1}{\it XMM-Newton} observation list.}
\small
\centering
\begin{tabular}{lllllll}
\hline
Source & ObsID & Obs. Date &  Exposure & Galactic &Distance & $N({\rm H}) $ \\
&&&(ks)&coordinates&(kpc)&$10^{21}$cm$^{-2}$ \\
\hline
  4U~1254--69		& 0060740101	& 22-01-2001	&$17.58$	&$(303.48;-6.42)$		&$ 13.0\pm 3.0 ^{a}$ 	&$2.15$	\\
			& 0060740901 	& 08-02-2002	&$29.62$	&			&			&	\\
			& 0405510301	& 13-09-2006	&$61.32$	&			&			&	\\
			& 0405510401	& 14-01-2007	&$62.91$	&			&			&	\\
			& 0405510501	& 09-03-2007	&$61.32$	&			&			&	\\
  4U~1543--62		& 0061140201	& 05-02-2001	&$50.10$	&$(321.76;-6.34)$		&$ 7.0 ^{b}$ &$2.46$	\\
  4U~1636--53		& 0500350301	& 29-09-2007	&$31.94$	&$(332.92;-4.82)$		&$ 6.0\pm 0.5 ^{c}$ 	&$2.64$	\\
			& 0500350401	& 27-02-2008	&$39.94$	&			&			&	\\
			& 0606070101	& 15-03-2009	&$41.18$	&			&			&	\\
			& 0606070301	& 05-09-2009	&$43.20$	&			&			&	\\
  4U~1735--44& 0090340201	& 03-09-2001  	&$21.77 $	&$(346.0;-6.9)$		&$9.4\pm 1.4 ^{d}$ 	&$ 2.56$	 \\ 
			& 0090340601 	&01-04-2013  &$ 85.00$	&			&			&	\\
  Cygnus~X--2		& 0111360101	&03-06-2002	&$21.54$	&$(87.33;-11.32)$ 	&$13.4\pm 2.0 ^{d}$ 	&$1.88$	\\
			& 0303280101	&14-06-2005	&$31.81$	&			&		&	\\
  GRO~J1655--40		& 0112921401	&14-03-2005	&$15.62$	&$(344.98;2.46)$		&$3.2\pm 0.2^{d}$	&$5.78$	\\
			& 0112921501	&15-03-2005	&$15.62$	&			&		&	\\
			& 0112921601	&16-03-2005	&$15.61$	&			&		&	\\
  GS~1826--238		& 0150390101	&08-04-2003	&$107.85$	&$(9.27;-6.09)$		&$6.7^{c}$	&$1.68$	\\
			& 0150390301	&09-04-2003	&$91.92$	&			&		&	\\
  GX~9+9			& 0090340101	&04-09-2001	&$20.06$	&$(8.51;9.04)$		&$4.4^{e}$	&$1.98$	\\
			& 0090340601	&25-09-2002	&$23.85$	&			&		&	\\
			& 0694860301	&28-03-2013	&$36.58$	&			&		&	\\
  GX~339--4		& 0148220201	&08-03-2003	&$20.50$	&$(338.94;-4.33)$		&$10.0\pm 4.5^{f}$	&$3.74$	\\
			& 0148220301	&20-03-2003	&$16.27$	&			&		&	\\
XTE~J1817--330 	&0311590501 &13-03-2006 &$20.73 $& $(359.8;-7.9)$ & $2.5\pm 1.5 ^{g}$ &$1.39$ \\   
\hline
\multicolumn{7}{p{14cm}|}{Hydrogen column density values are obtained from \citet{kal05}. (a)\citet{int03}; (b) \citet{wan04}; (c) \citet{gal06}; (d)  \citet{jon04}; (e) \citet{gri02}; (f) \citet{hyn04}; (g) \citet{sal06}. }
\end{tabular}
\end{table*}
 
\section{Observations and data reduction}\label{sec_dat}
We have analyzed 10 bright LMXBs spectra obtained with the {\it XMM-Newton}
observatory. {\it XMM-Newton} carries two high spectral resolution instruments,
the Reflection Grating Spectrometers \citep[RGS;][]{den01}. Each RGS consists
of an array of reflection gratings which allows the diffraction of the X-rays,
which are then detected on the charge couple devices (CCDs). The maximum
instrumental resolution is $\Delta\lambda\sim$ 0.06\AA\, with a maximum
effective area of about 140~cm$^{2}$ around 15~\AA. The pileup effect, the
detection of two events simultaneously as one single event, does not affect the
O K-edge absorption region (21--24\AA), and thus it can be ignored.
Table~\ref{tab1} shows the specifications of the sources analyzed, including
hydrogen column density 21~cm measurements from the \citet{kal05} survey, while
Figure~\ref{fig1b} shows the location in Galactic coordinates of all these
source.  We have reduced the data with the Science Analysis System (SAS,
version 14.0.0) using the standard procedure to obtain the RGS spectra. A total
of 25 observations were analyzed. We use $\chi^{2}$ statistic with the
weighting method for low counts regime defined by \citet{chu96}. The spectral
fitting was performed with the {\sc xspec} analysis data package \citep[version
12.9.0e\footnote{\url{https://heasarc.gsfc.nasa.gov/xanadu/xspec/}}]{arn96}
%

   \begin{figure}
   \begin{center}
     \includegraphics[scale=0.6]{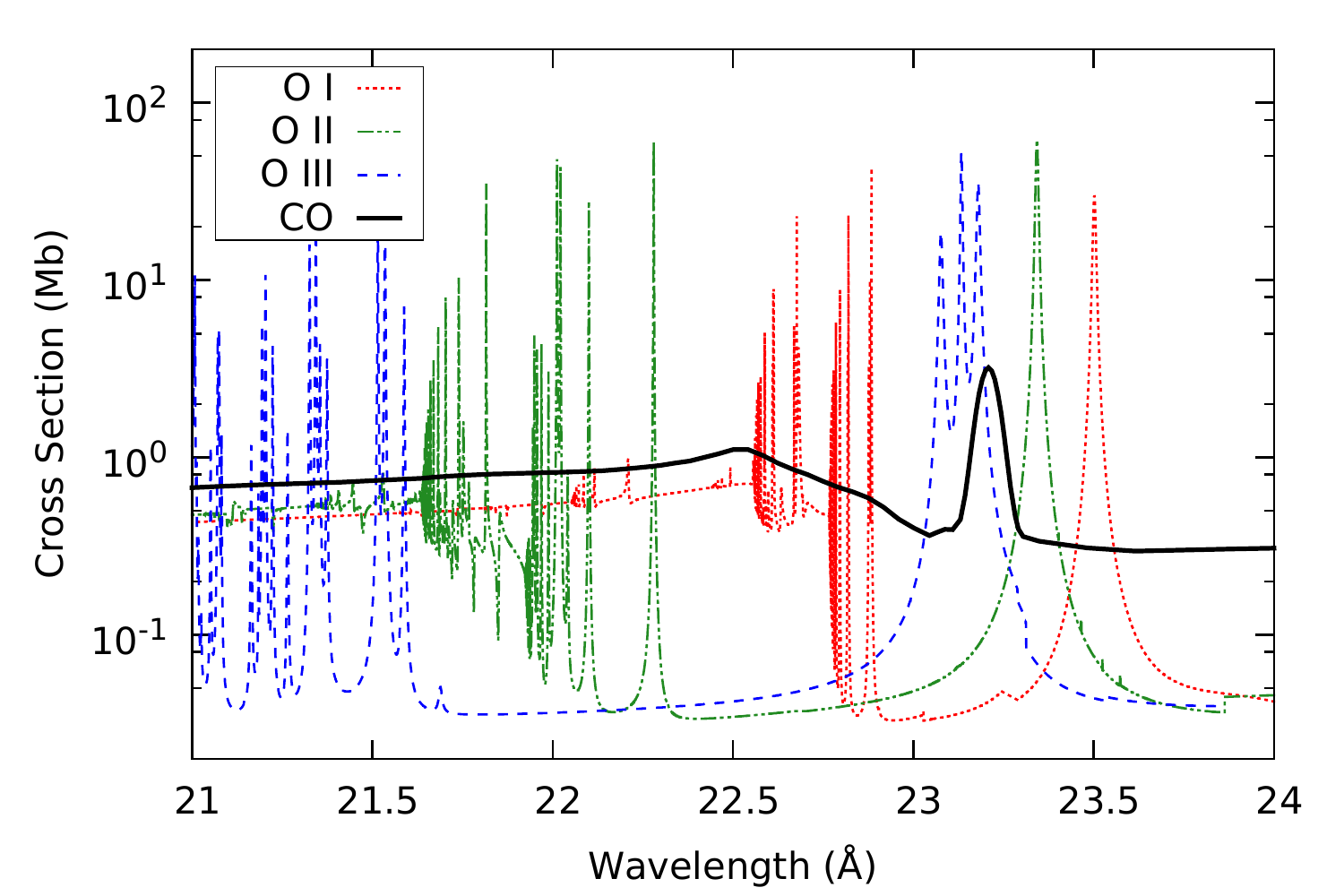}
     \caption{Photoionization cross-sections included on {\tt ISMabs} for the O
     K-edge modeling. O~{\sc I} was computed by \citet{gor13}, while O~{\sc II} and
     O~{\sc III} were computed by \citet{gar05}. The CO cross-section was measured
     experimentally by \citet{bar79}.}\label{fig1}
        \end{center}
   \end{figure}

   \begin{figure}
   \begin{center}
     \includegraphics[scale=0.37]{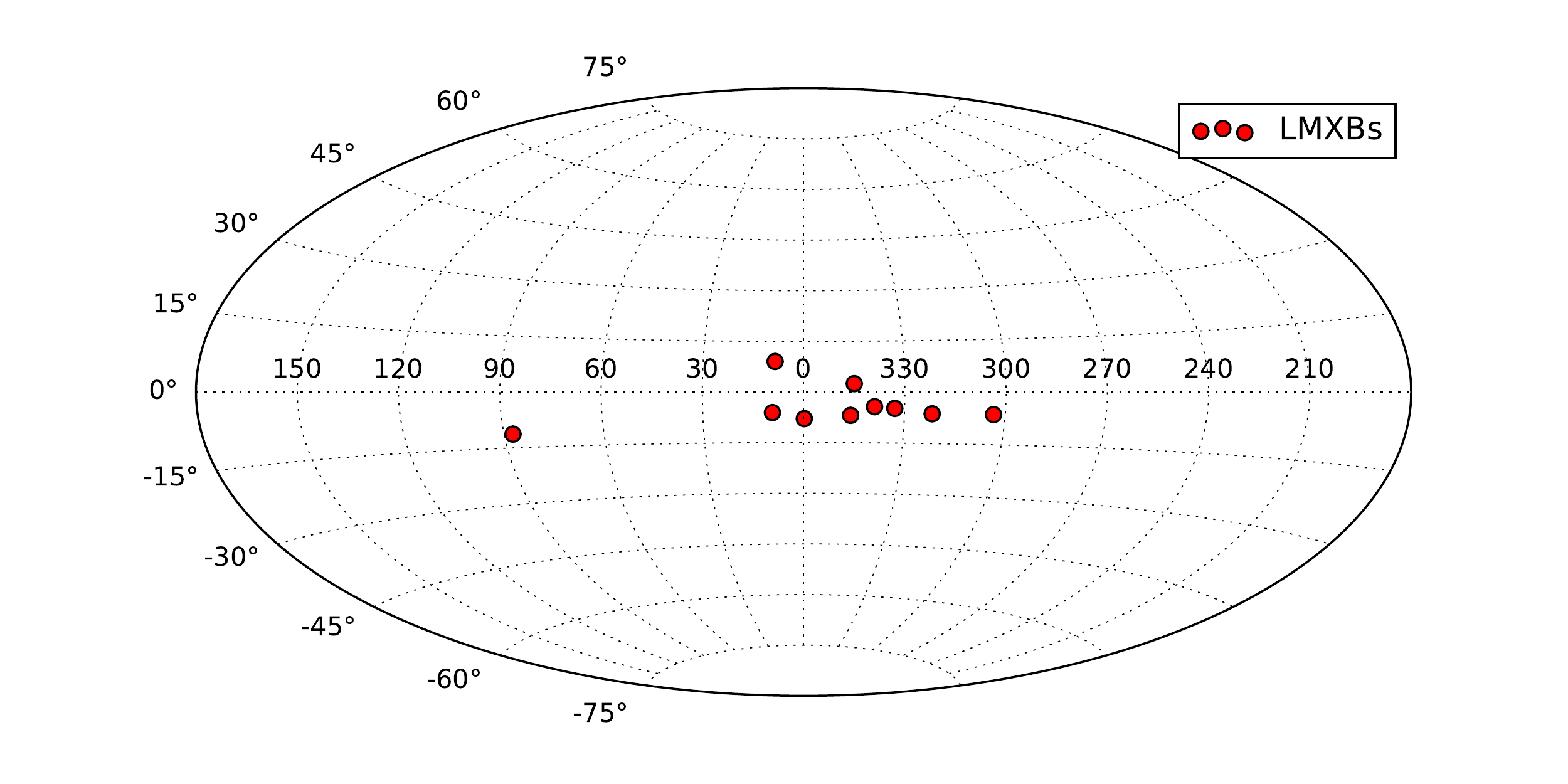}
     \caption{Aitoff projection of the location for all low mass X-ray binaries
     analyzed in this work.}\label{fig1b}
        \end{center}
   \end{figure}

\section{O K-edge modeling}\label{sec_ofit}
In order to model the O K-edge absorption region (21--24 \AA) we used a simple
{\tt power-law} model for the continuum and the {\tt ISMabs} X-ray absorption
model which includes neutrally, singly, and doubly ionized species of H, He,
 C, N, O, Ne, Mg, Si, S, Ar and Ca \citep{gat15}. We fixed the H column
densities to the $N({\rm H})$ values reported by \citet{gat16a}, which were
obtained through a broadband fit (11--24 \AA) for all sources included in the
present work. In the case of 4U~1543--62, which was not included in the
\citet{gat16a} analysis, we use the $N({\rm H})$ value from the 21~$cm$ measurements indicated in
Table~\ref{tab1}. For each source, we fitted all observation simultaneously
using the same {\tt Photon-index} as well as the column densities for O~{\sc
  I}, O~{\sc II} and O~{\sc III}, while allowing the normalization to vary.
This accounts for the possibility of variability in the flux from the LMXB while
maintaining the ISM absorption at a fixed value.
%

\begin{table*}
\small
\caption{\label{tab2}O K-edge fit results.}
\centering
\begin{tabular}{lcccccccccc}
\hline
Source&  $N({\rm H})$ &  $N({\rm O~{\sc I}})$& $N({\rm O~{\sc II}})$& $N({\rm O~{\sc III}})$&  $\chi^2$/dof & $\Delta \chi^2$ & $N({\rm CO})$\\
& ($10^{21}$cm$^{-2}$) & ($10^{17}$cm$^{-2}$)& ($10^{16}$cm$^{-2}$)& ($10^{16}$cm$^{-2}$)&   & With CO & ($10^{16}$cm$^{-2}$)  \\
			\hline
  4U~1254--69& $2.2$		&$16.82  \pm 14.29 $		&$ 3.67 \pm 2.62 $	&$  2.47 \pm 2.06 $			&$1918/1762$&$<8$& --\\  
  4U~1543--62&	$2.46$	&$19.45  \pm  2.04$		&$6.38  \pm 4.27 $	&$ 4.23 \pm 3.67  $			&$275/291$&$<4$& --\\
  4U~1636--53&$5.6$		&$24.92  \pm  1.35$		&$ 8.73 \pm 3.34 $	&$ 7.38 \pm 4.05 $				&$1412/1181$&$<12$ & $7.08 \pm 3.45$\\
  4U~1735-44&	$3.2$		&$19.98  \pm  2.35$		&$6.64  \pm  2.27$	&$ 5.52 \pm 3.96 $			&$657/583$&$<7$& --\\
  Cygnus~X--2& $4.3$		&$15.05  \pm 0.68 $		&$6.65  \pm 0.76 $	&$3.59  \pm2.18  $			&$709/588$&$<9$& --\\
  GRO~J1655--40	& $7.8$	&$ 32.47 \pm 2.30 $		&$ 5.51 \pm 2.87 $	&$ 3.45 \pm 2.52  $			&$1042/882$&$<1$& --\\\
  GS~1826--238&	$3.1$	&$ 26.27 \pm 1.99 $		&$5.85  \pm  3.10$	&$ 0.07\pm 0.05$		&$671/590$&$<1$& --\\
  GX~9+9	&$7.4$			&$18.73  \pm 2.04 $		&$ 4.63 \pm 4.29 $	&$ 3.92 \pm 3.57 $			&$330/589$&$<3$& --\\
  GX~339--4&	 $4.1$		&$32.03  \pm 2.11 $		&$ 8.79 \pm 4.25 $	&$6.14  \pm 3.04 $			&$671/592$&$<1$& --\\
  XTE~J1817-330&$1.4$	&$13.18  \pm 1.45 $		&$ 8.34 \pm  3.69$		&$4.50  \pm 1.97 $	&$385 /293$&$< 12$ & $ 7.22 \pm 0.57$\\
 \hline
\end{tabular}
\end{table*}

The use of accurate atomic data is crucial to model the O K-edge absorption
features (21--24\AA). Owing to relaxation effects (Auger damping) the
absorption K-edge does not have a simple edge shape but instead shows multiple
resonances which lead to the smearing of the edge when viewed with an instrument
with finite resolution \citep{gar05}. {\tt ISMabs}
includes O~{\sc I} cross-section from \citet{gor13} and O~{\sc II}, O~{\sc III}
cross-sections from \citet{gar05}. These cross-sections have been improved by
using astrophysical observations as reference for the energy position of the
resonances, making them the best atomic data currently available for
high-resolution X-ray spectroscopy analysis \citep{gat14,gat15}.
%

   \begin{figure*}
   \begin{center}
     \includegraphics[width=\linewidth]{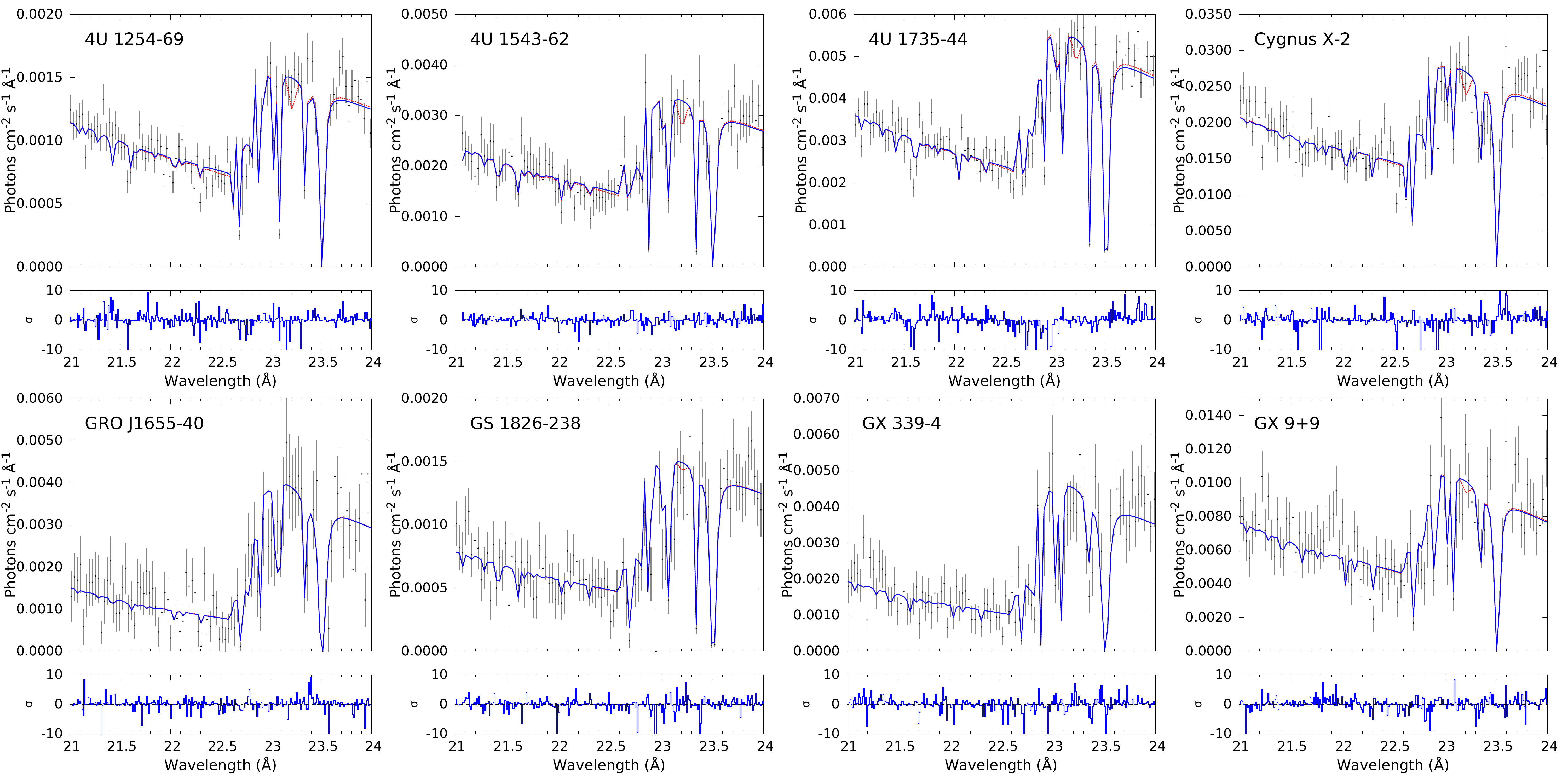}
     \caption{Oxygen K-edge region {\tt ISMabs} best fits without CO (blue dotted lines) and including CO (red solid lines). For all sources above we can safely reject a
successful detection of CO (see text for details).}\label{fig3a}
     \end{center}
   \end{figure*}

   \begin{figure*}
   \begin{center}
     \includegraphics[scale=0.4]{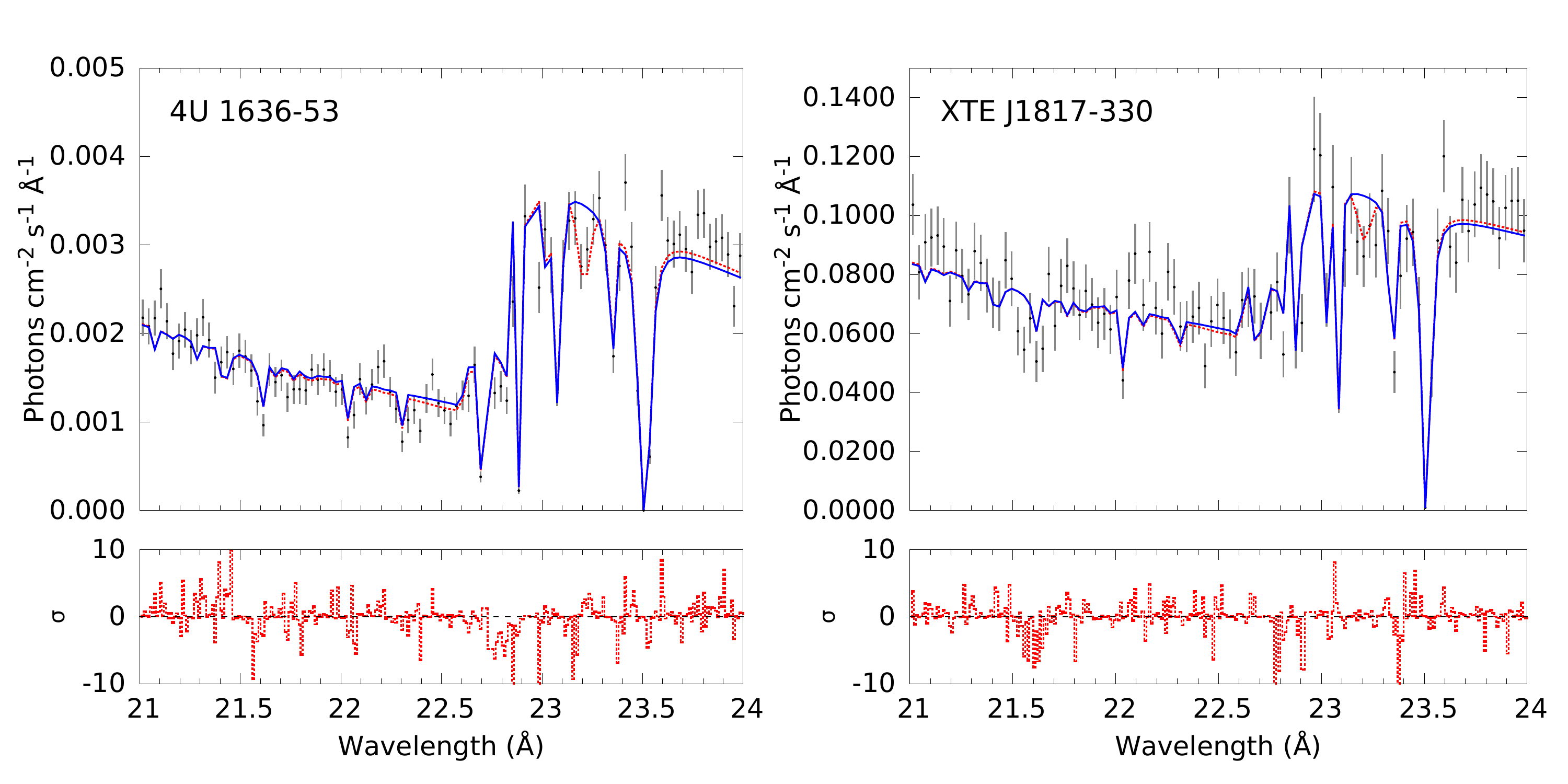}
     \caption{{\tt ISMabs} best fit of the oxygen K-edge region without CO (blue
     solid lines) and including CO (red dotted lines). For both sources we obtained a significant CO detection
with at least $95\%$ confidence.}\label{fig3b}
     \end{center}
   \end{figure*} 
   
After the best fit is found by using only an atomic component, we allow the CO
column density to vary. In order to model the CO molecular absorption, we use
the cross-section measured experimentally by \citet{bar79}. This cross section
does not includes the measurements at the resonance. Instead, \cite{bar79}
estimated the peak intensities of the resonance based on measurements with
different column densities (see their Table~4) and using these values we can
obtain a complete cross section by joining them smoothly. A complete CO photoabsorption cross-section has been incorporated in the {\sc spex} data analysis package\footnote{\url{http://var.sron.nl/SPEX-doc/cookbookv3.0/cookbook.html}} and we have made use of it.
Figure~\ref{fig1}
shows the cross-sections included on {\tt ISMabs} in order to model the O
K-edge. The principal CO resonance, located at $\sim$23.20 \AA, corresponds to
the excitation to the 2$p\pi^{*}$ orbital and $^{1}\Pi$ state with a high
width, compared with the O~{\sc I}, O~{\sc II} and O~{\sc III} resonances, attributed to the vibrational level excitation. In that respect, \citet{bar79} quoted an uncertainty in the energy scale for the measurements of $\pm$ $0.3$ eV ($\pm$ 0.026 \AA) while theoretical analysis of molecular vibrational modes indicate a vibrational spacing $<0.14$ eV ($<0.012$ \AA) \citep{dom90}.  Finally, this strong
feature is partially embedded in the O~{\sc III} K$\alpha$ triplet, making
difficult its detection.

    \begin{figure*}
     \includegraphics[scale=0.25]{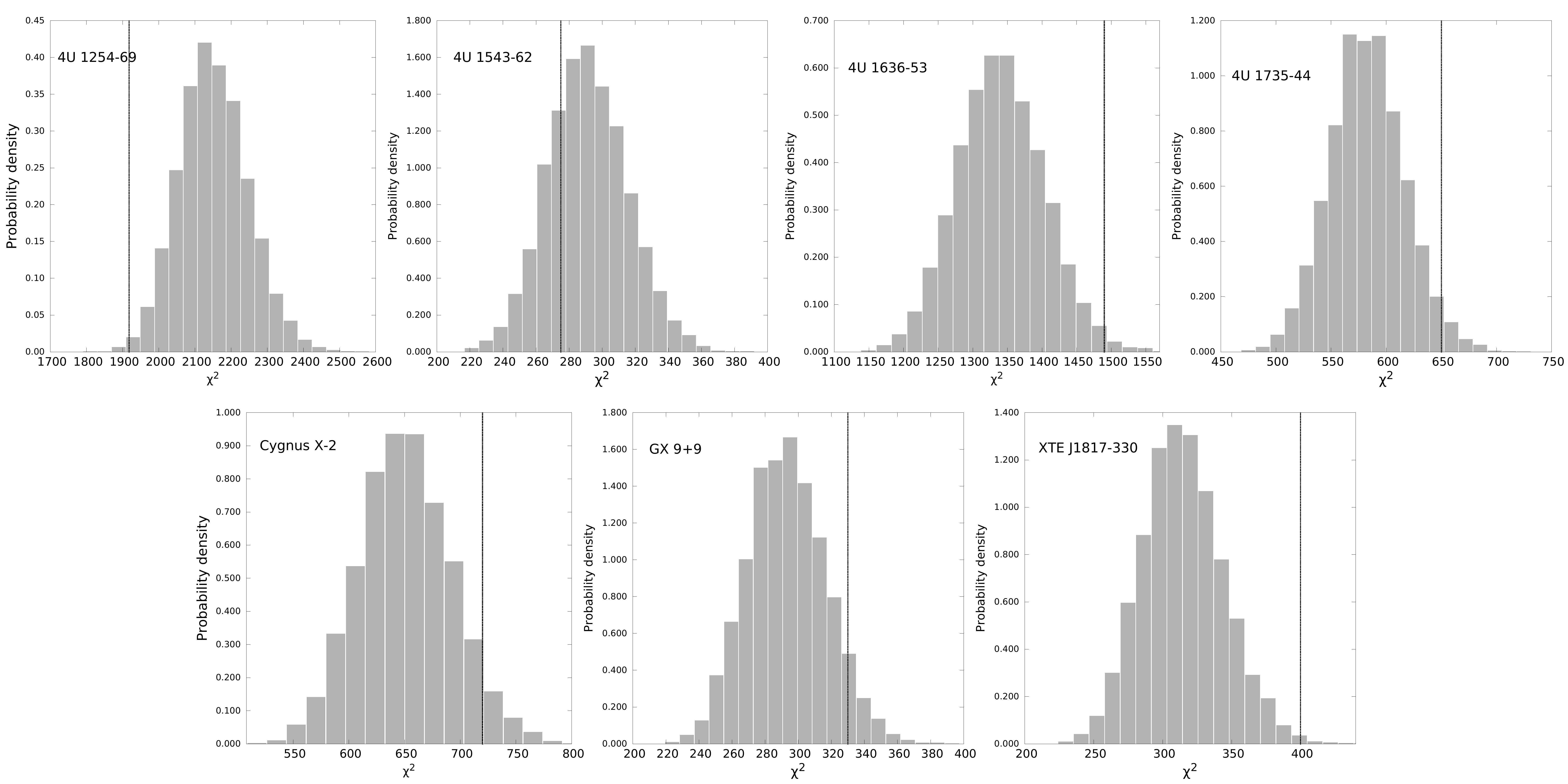}
     \caption{Monte-Carlo simulation analysis to estimate the
     significance of the detection of the CO absorption feature. Vertical dashed lines
     correspond to the best $\chi^{2}$ obtained for each source. }\label{fig4}
   \end{figure*}

\section{Results and discussions}\label{sec_dis}

Table \ref{tab2} shows the results from the best fits for all sources analyzed.
As was detailed on Section~\ref{sec_ofit}, first we performed a fit allowing
the O~{\sc I}, O~{\sc II} and O~{\sc III} column densities to vary as well as
the {\tt powerlaw} parameters. After obtain the best fit, we allow the CO
column density to vary, as well as the O column densities, until we obtain the
best fit.  $\Delta \chi^{2}$ values on Table~\ref{tab2} correspond to the
statistical difference, in units of $\chi^{2}$, between the model without CO
and the model including CO. The largest increases in the statistic correspond to
the objects 4U~1636-53 and XTE~J1817-330 (both with $\Delta \chi^{2}<12$).

Figures~\ref{fig3a} and~\ref{fig3b} show the best fit of our model to the O
K-edge region for all the sources selected. Observations for the same source
are combined for illustrative purposes only. Figure~\ref{fig3a} shows those
sources for which the CO absorption feature has not been statistically
detected, while Figure~\ref{fig3b} shows the two sources for which a successful
CO detection has been performed (see explanation below).  In general, the
residuals obtained without the CO contribution are small and evenly distributed
along the wavelength region, indicating satisfactory modeling only with an atomic
model. This is reflected in the fit statistics which correspond, in all cases,
to $\chi^{2}/dof\sim 1$ (see Table~\ref{tab2}). For sources in which the
inclusion of CO yielded to some improvement in the fit statistics, the presence
of the CO resonance at $23.20$\AA\ can be appreciated in the models shown in
Figures~\ref{fig3a} and~\ref{fig3b}.  However, in most cases the statistical
improvement --referring to the variation on the $\chi^{2}$ value-- is low.

We have performed a Monte-Carlo (MC) simulation analysis of the goodness
of fit in order to estimate the significance of the CO detection for all
sources with $\Delta \chi^{2}>2$. First, we fit the spectra with the reference
model (i.e. {\tt ISMabs}* {\tt powerlaw}, without including CO), obtaining a
$\chi^{2}$ statistical value. Second, we allow the CO component to vary,
obtaining  a second $\chi^{2}$ statistical value. This procedure was performed $10^{4}$
times by generating synthetic spectra for the source and background with the same
exposure times as the combined data. Each fake spectrum (i.e. without CO) was fitted with a model which include CO as free parameter and the $\chi^{2}$ value
was recorded. Following this procedure, the number of simulations where the
reduction in the $\chi^{2}$ is at least as large as the observed in the real
fit relative to the total number of simulations gives the probability of a
false detection.

Figure~\ref{fig4} shows the MC results for the sources with $\Delta
\chi^{2}>2$. Vertical dashed lines correspond to the $\chi^{2}$ value for the
best fit model including CO. Table~\ref{tab3} shows false detection probability
values derived from the MC simulations. We computed the significance of the CO
possible detection in units of $\sigma$. Finally, the minimum $\sigma$ units
required to ensure a $95\%$ significant detection is included for each case.
This last value depends on the degrees of freedom involved, i.e., the number of
free parameters in the O K-edge fit, which are also indicated in
Table~\ref{tab3}.

\begin{table}
\small
\caption{\label{tab3}Monte-Carlo analysis.}
\begin{tabular}{lcccc}
\hline
Source& d.o.f. &  $\rho ^{a}$ &  $\sigma^{b}$  &Significant\\
	&	 & 	 &  	 							 &Regime$^{c}$\\
\hline
  4U~1254--69& 3 	&	$99.69$ 		& $<0.35$ &$3.84$\\  
  4U~1543--62&	8&$78.52$  		&$<4.59$ & $15.51 $ \\  
  4U~1735-44&4	&$13.46$  		& $<7.78$& $9.49$\\  
  Cygnus~X--2&	4&$10.87 $  		& $<7.78$& $9.49 $\\  
  GX~9+9	&5	&  $10.08$		& $<9.24$& $11.07 $\\  
  4U~1636--53&6	&$3$  		&$<6.64$ & $3.84  $\\  
  XTE~J1817-330&	3	&  $2$		& $<6.64$& $3.84 $\\  
 \hline
 \multicolumn{5}{p{8cm}|}{(a) False detection probability (in $\%$); (b) Significance of the
possible CO detection (in $\sigma$ units); (c) Minimum $\sigma$ required to
ensure a $95\%$ significant detection. }
\end{tabular}
 \end{table}

For 4U~1254--69, 4U~1543--62, 4U~1735-44, Cygnus~X--2, and GX~9+9 we obtained a
false detection probability of $> 10 \%$, and therefore we can safely reject a
successful detection of CO toward these lines-of-sight. On the other hand, we
obtained a false detection probability of $3\%$ and $2\%$ in 4U~1636--53 and
XTE~J1817-330, respectively. These values correspond to a significant detection
with at least $95\%$ confidence. The column density value measured are $N({\rm
CO})= (7.22 \pm 0.57)$ $\times 10^{16}$cm$^{-2}$ (for XTE~J1817-330) and
$N({\rm CO})= (7.08 \pm 3.45)$ $\times 10^{16}$cm$^{-2}$ (for 4U~1636--53).
%

\section{Comparison with Previous Work}\label{sec_previous}
\citet{pin10} modeled the LMXB GS~1826--238 {\it XMM-Newton} high-resolution
spectra using the {\sc spex} analysis software, which includes the molecular
absorption model ({\tt amol}). They found an upper limit for the CO column
density of $< 4.0\times10^{16}$cm$^{-2}$, as well as upper limits for
Ca$_{3}$Fe$_{2}$Si$_{3}$O$_{12}$, amorphous ice (H$_{2}$O), carbon monoxide
(CO), and hercynite (FeAl$_{2}$O$_{4}$) column densities. \citet{pin13} also
report the detection of CO analyzing nine LMXBs, six of which are included in
this work. The CO column density values vary between $0.02-0.7\times10^{17}$~cm$^{-2}$
along the multiple lines-of-sight. In the case of 4U~1636--53, they estimate a
column density of $N({\rm CO})= (3.0 \pm 2.0)$ $\times 10^{16}$ cm$^{-2}$.
However, the O~{\sc I}, O~{\sc II}, and O~{\sc III} atomic cross sections
incorporated in the {\sc spex} models do not include the effects of Auger
damping, which has a significant effect in the structure of the K-edge.
This limits the ability of these models to reproduce the observations.

On the other hand, our findings are consistent with those of
\citet{gat13a,gat13b}, who achieved good O K-edge fits to {\it Chandra}
high-resolution spectra using a {\tt warmabs} model based on the {\sc xstar}
photoionization code, without clear evidence for the presence of CO X-ray
absorption features. More recently, \citet{gat16a}  have performed a
comprehensive analysis of the ISM along 24 lines of sight using both {\it
Chandra} and {\it XMM-Newton} high-resolution spectra, demonstrating statistically
acceptable fits without molecular or solid contribution to the absorption in the O K-edge wavelength
region.

Large-scale CO surveys show that CO is the dominant molecule in the interstellar medium, after
  $H_2$, and that its abundance relative to $H_2$ is $\sim 10^{-4}$, i.e. most
  oxygen is bound in the molecule in the cold phase of the ISM \citep{her73}.  The major contribution of CO in the Milky Way
is expected to be in the Galactic plane, with higher CO column densities around
the galactic center \citep{bit97,dam01,nak06,pad06,pet15}. However, the
Galactic extinction toward these lines-of-sight
makes it difficult to obtain spectra which are well suited to the study of the O K-edge
region. Therefore, the sources included in our sample satisfy (1) line-of-sight
toward or near the galactic plane, and (2) high counts rate in the 21--24~\AA\
wavelength region, allowing the modeling not only of O~{\sc I} K$\alpha$, but
also O~{\sc II} and O~{\sc III} K$\alpha$ lines. However, according to
the survey of \citet{dam01}, the presence of CO in the line of sight toward our sample is
predicted to be low, with an average value of  $\sim 3.5\times 10^{16}$ cm$^{-2}$ for $|l|<10$. 
Using {\it XMM-Newton} response files we simulated spectra assuming
typical O~{\sc I}, O~{\sc II}, and O~{\sc III} column densities measured by
\citet{gat16a}, and a typical exposure time (e.g. 30 ks) in order to determine
the minimum CO column density required to perform a successful detection. We
found a lower limit value corresponding to $N({\rm CO})>6$ $\times 10^{16}$
cm$^{-2}$. In the case of {\it Chandra} we found a lower limit value to perform
a significant detection of $N({\rm CO})>12$ $\times 10^{16}$
cm$^{-2}$. These values are comparable to the upper limits and detections that
  we report in the previous section.

The presence of CO along the line-of-sight toward XTE--J1817-330 has not been
reported before. XTE--J1817-330 constitutes a bright source with high
signal-to-noise ratio in the O K-edge wavelength region to allow the
identification of K$\beta$ and K$\gamma$ O~{\sc I} absorption lines
\citep{gat13a,gat13b}. The quality of the spectra is reflected in the error bar
of the CO column density, which correspond to a $7 \%$ of the measured value.
In the case of 4U~1636--53, we agree with CO column density estimated by
\citet{pin13}. However, it should be noted that the error for the CO column
density in our modeling is large enough that our measurement 
should be taken as an upper limit. 
%


\section{Conclusions}\label{sec_con}
We reported on the spectral analysis of 10 Galactic X-ray binaries located in
the Galactic plane and the search for CO X-ray absorption features in these
spectra. A total of 25 {\it XMM-Newton} observations were analyzed. We modeled
the O K-edge absorption region (21--24~\AA) using the {\tt ISMabs} X-ray
absorption model, which includes singly, and doubly ionized photoabsorption
cross-sections for O. For all sources we were able to asses the presence and
strength of O~{\sc I}, O~{\sc II}, and O~{\sc III} absorption features,
measuring the column density for each ion. We include the CO experimental
photo-absorption cross section measured by \citet{bar79} in order to model the
contribution of CO to the absorption spectra. We performed Monte-Carlo
simulations to obtain a rigorous estimate of the statistical significance of
possible CO detection concluding that the statistical analysis prevents from a
significant detection of CO molecular X-ray absorption features in 8 of the
sources analyzed. Finally, we measured CO column density values for
XTE--J1817-330 ($7.22 \pm 0.57$ $\times 10^{16}$ cm$^{-2}$), and 4U~1636-53
($7.08 \pm 3.45$ $\times 10^{16}$ cm$^{-2}$). This, along with simulations, suggests
that the statistical quality of the current archive of XMM-Newton RGS observations is generally
not sufficient to detect CO from a large fraction of the sight lines which have been observed.
The detections likely represent sight lines with columns greater than average.
Deeper observations with the existing instruments have the potential to detect CO along more lines of sight.

\bibliographystyle{mnras}

\begin{thebibliography}{}
\makeatletter
\relax
\def\mn@urlcharsother{\let\do\@makeother \do\$\do\&\do\#\do\^\do\_\do\%\do\~}
\def\mn@doi{\begingroup\mn@urlcharsother \@ifnextchar [ {\mn@doi@}
  {\mn@doi@[]}}
\def\mn@doi@[#1]#2{\def\@tempa{#1}\ifx\@tempa\@empty \href
  {http://dx.doi.org/#2} {doi:#2}\else \href {http://dx.doi.org/#2} {#1}\fi
  \endgroup}
\def\mn@eprint#1#2{\mn@eprint@#1:#2::\@nil}
\def\mn@eprint@arXiv#1{\href {http://arxiv.org/abs/#1} {{\tt arXiv:#1}}}
\def\mn@eprint@dblp#1{\href {http://dblp.uni-trier.de/rec/bibtex/#1.xml}
  {dblp:#1}}
\def\mn@eprint@#1:#2:#3:#4\@nil{\def\@tempa {#1}\def\@tempb {#2}\def\@tempc
  {#3}\ifx \@tempc \@empty \let \@tempc \@tempb \let \@tempb \@tempa \fi \ifx
  \@tempb \@empty \def\@tempb {arXiv}\fi \@ifundefined
  {mn@eprint@\@tempb}{\@tempb:\@tempc}{\expandafter \expandafter \csname
  mn@eprint@\@tempb\endcsname \expandafter{\@tempc}}}

\bibitem[\protect\citeauthoryear{{Arnaud}}{{Arnaud}}{1996}]{arn96}
{Arnaud} K.~A.,  1996, in {Jacoby} G.~H.,  {Barnes} J.,  eds,  Astronomical
  Society of the Pacific Conference Series Vol. 101, Astronomical Data Analysis
  Software and Systems V. p.~17

\bibitem[\protect\citeauthoryear{{Barrus}, {Blake}, {Burek}, {Chambers}  \&
  {Pregenzer}}{{Barrus} et~al.}{1979}]{bar79}
{Barrus} D.~M.,  {Blake} R.~L.,  {Burek} A.~J.,  {Chambers} K.~C.,
  {Pregenzer} A.~L.,  1979, \mn@doi [\pra] {10.1103/PhysRevA.20.1045}, \href
  {http://adsabs.harvard.edu/abs/1979PhRvA..20.1045B} {20, 1045}

\bibitem[\protect\citeauthoryear{{Bitran}, {Alvarez}, {Bronfman}, {May}  \&
  {Thaddeus}}{{Bitran} et~al.}{1997}]{bit97}
{Bitran} M.,  {Alvarez} H.,  {Bronfman} L.,  {May} J.,   {Thaddeus} P.,  1997,
  \mn@doi [\aaps] {10.1051/aas:1997214}, \href
  {http://adsabs.harvard.edu/abs/1997A%26AS..125...99B} {125}

\bibitem[\protect\citeauthoryear{{Churazov}, {Gilfanov}, {Forman}  \&
  {Jones}}{{Churazov} et~al.}{1996}]{chu96}
{Churazov} E.,  {Gilfanov} M.,  {Forman} W.,   {Jones} C.,  1996, \mn@doi
  [\apj] {10.1086/177997}, \href
  {http://adsabs.harvard.edu/abs/1996ApJ...471..673C} {471, 673}

\bibitem[\protect\citeauthoryear{{Dame}, {Hartmann}  \& {Thaddeus}}{{Dame}
  et~al.}{2001}]{dam01}
{Dame} T.~M.,  {Hartmann} D.,   {Thaddeus} P.,  2001, \mn@doi [\apj]
  {10.1086/318388}, \href {http://adsabs.harvard.edu/abs/2001ApJ...547..792D}
  {547, 792}

\bibitem[\protect\citeauthoryear{{Domke}, {Xue}, {Puschmann}, {Mandel},
  {Hudson}, {Shirley}  \& {Kaindl}}{{Domke} et~al.}{1990}]{dom90}
{Domke} M.,  {Xue} C.,  {Puschmann} A.,  {Mandel} T.,  {Hudson} E.,  {Shirley}
  D.~A.,   {Kaindl} G.,  1990, \mn@doi [Chemical Physics Letters]
  {10.1016/0009-2614(90)85314-3}, \href
  {http://adsabs.harvard.edu/abs/1990CPL...173..122D} {173, 122}

\bibitem[\protect\citeauthoryear{{Galloway}, {Psaltis}, {Muno}  \&
  {Chakrabarty}}{{Galloway} et~al.}{2006}]{gal06}
{Galloway} D.~K.,  {Psaltis} D.,  {Muno} M.~P.,   {Chakrabarty} D.,  2006,
  \mn@doi [The Astrophysical Journal Letters] {10.1086/499579}, \href
  {http://cdsads.u-strasbg.fr/abs/2006ApJ...639.1033G} {639, 1033}

\bibitem[\protect\citeauthoryear{{Garc{\'{\i}}a}, {Mendoza}, {Bautista},
  {Gorczyca}, {Kallman}  \& {Palmeri}}{{Garc{\'{\i}}a} et~al.}{2005}]{gar05}
{Garc{\'{\i}}a} J.,  {Mendoza} C.,  {Bautista} M.~A.,  {Gorczyca} T.~W.,
  {Kallman} T.~R.,   {Palmeri} P.,  2005, \mn@doi [\apjs] {10.1086/428712},
  \href {http://adsabs.harvard.edu/abs/2005ApJS..158...68G} {158, 68}

\bibitem[\protect\citeauthoryear{{Gatuzz} et~al.,}{{Gatuzz}
  et~al.}{2013a}]{gat13a}
{Gatuzz} E.,  et~al., 2013a, \mn@doi [\apj] {10.1088/0004-637X/768/1/60}, \href
  {http://adsabs.harvard.edu/abs/2013ApJ...768...60G} {768, 60}

\bibitem[\protect\citeauthoryear{{Gatuzz} et~al.,}{{Gatuzz}
  et~al.}{2013b}]{gat13b}
{Gatuzz} E.,  et~al., 2013b, \mn@doi [\apj] {10.1088/0004-637X/778/1/83}, \href
  {http://adsabs.harvard.edu/abs/2013ApJ...778...83G} {778, 83}

\bibitem[\protect\citeauthoryear{{Gatuzz}, {Garc{\'{\i}}a}, {Mendoza},
  {Kallman}, {Bautista}  \& {Gorczyca}}{{Gatuzz} et~al.}{2014}]{gat14}
{Gatuzz} E.,  {Garc{\'{\i}}a} J.,  {Mendoza} C.,  {Kallman} T.~R.,  {Bautista}
  M.~A.,   {Gorczyca} T.~W.,  2014, \mn@doi [\apj]
  {10.1088/0004-637X/790/2/131}, \href
  {http://adsabs.harvard.edu/abs/2014ApJ...790..131G} {790, 131}

\bibitem[\protect\citeauthoryear{{Gatuzz}, {Garc{\'{\i}}a}, {Kallman},
  {Mendoza}  \& {Gorczyca}}{{Gatuzz} et~al.}{2015}]{gat15}
{Gatuzz} E.,  {Garc{\'{\i}}a} J.,  {Kallman} T.~R.,  {Mendoza} C.,   {Gorczyca}
  T.~W.,  2015, \mn@doi [\apj] {10.1088/0004-637X/800/1/29}, \href
  {http://adsabs.harvard.edu/abs/2015ApJ...800...29G} {800, 29}

\bibitem[\protect\citeauthoryear{{Gatuzz}, {Garc{\'{\i}}a}, {Kallman}  \&
  {Mendoza}}{{Gatuzz} et~al.}{2016}]{gat16a}
{Gatuzz} E.,  {Garc{\'{\i}}a} J.~A.,  {Kallman} T.~R.,   {Mendoza} C.,  2016,
  preprint, \href {http://adsabs.harvard.edu/abs/2016arXiv160206955G} {}
  (\mn@eprint {arXiv} {1602.06955})

\bibitem[\protect\citeauthoryear{{Gorczyca} et~al.,}{{Gorczyca}
  et~al.}{2013}]{gor13}
{Gorczyca} T.~W.,  et~al., 2013, \mn@doi [\apj] {10.1088/0004-637X/779/1/78},
  \href {http://adsabs.harvard.edu/abs/2013ApJ...779...78G} {779, 78}

\bibitem[\protect\citeauthoryear{{Grimm}, {Gilfanov}  \& {Sunyaev}}{{Grimm}
  et~al.}{2002}]{gri02}
{Grimm} H.-J.,  {Gilfanov} M.,   {Sunyaev} R.,  2002, \mn@doi [\aap]
  {10.1051/0004-6361:20020826}, \href
  {http://cdsads.u-strasbg.fr/abs/2002A%26A...391..923G} {391, 923}

\bibitem[\protect\citeauthoryear{{Herbst} \& {Klemperer}}{{Herbst} \&
  {Klemperer}}{1973}]{her73}
{Herbst} E.,  {Klemperer} W.,  1973, \mn@doi [\apj] {10.1086/152436}, \href
  {http://adsabs.harvard.edu/abs/1973ApJ...185..505H} {185, 505}

\bibitem[\protect\citeauthoryear{{Hynes}, {Steeghs}, {Casares}, {Charles}  \&
  {O'Brien}}{{Hynes} et~al.}{2004}]{hyn04}
{Hynes} R.~I.,  {Steeghs} D.,  {Casares} J.,  {Charles} P.~A.,   {O'Brien} K.,
  2004, \mn@doi [\apj] {10.1086/421014}, \href
  {http://cdsads.u-strasbg.fr/abs/2004ApJ...609..317H} {609, 317}

\bibitem[\protect\citeauthoryear{{Jonker} \& {Nelemans}}{{Jonker} \&
  {Nelemans}}{2004}]{jon04}
{Jonker} P.~G.,  {Nelemans} G.,  2004, \mn@doi [\mnras]
  {10.1111/j.1365-2966.2004.08193.x}, \href
  {http://cdsads.u-strasbg.fr/abs/2004MNRAS.354..355J} {354, 355}

\bibitem[\protect\citeauthoryear{{Juett}, {Schulz}  \& {Chakrabarty}}{{Juett}
  et~al.}{2004}]{jue04}
{Juett} A.~M.,  {Schulz} N.~S.,   {Chakrabarty} D.,  2004, \mn@doi [\apj]
  {10.1086/422511}, \href {http://adsabs.harvard.edu/abs/2004ApJ...612..308J}
  {612, 308}

\bibitem[\protect\citeauthoryear{{Juett}, {Schulz}, {Chakrabarty}  \&
  {Gorczyca}}{{Juett} et~al.}{2006}]{jue06}
{Juett} A.~M.,  {Schulz} N.~S.,  {Chakrabarty} D.,   {Gorczyca} T.~W.,  2006,
  \mn@doi [\apj] {10.1086/506189}, \href
  {http://adsabs.harvard.edu/abs/2006ApJ...648.1066J} {648, 1066}

\bibitem[\protect\citeauthoryear{{Kalberla}, {Burton}, {Hartmann}, {Arnal},
  {Bajaja}, {Morras}  \& {P{\"o}ppel}}{{Kalberla} et~al.}{2005}]{kal05}
{Kalberla} P.~M.~W.,  {Burton} W.~B.,  {Hartmann} D.,  {Arnal} E.~M.,  {Bajaja}
  E.,  {Morras} R.,   {P{\"o}ppel} W.~G.~L.,  2005, \mn@doi [\aap]
  {10.1051/0004-6361:20041864}, \href
  {http://adsabs.harvard.edu/abs/2005A%26A...440..775K} {440, 775}

\bibitem[\protect\citeauthoryear{{Liao}, {Zhang}  \& {Yao}}{{Liao}
  et~al.}{2013}]{lia13}
{Liao} J.-Y.,  {Zhang} S.-N.,   {Yao} Y.,  2013, \mn@doi [\apj]
  {10.1088/0004-637X/774/2/116}, \href
  {http://adsabs.harvard.edu/abs/2013ApJ...774..116L} {774, 116}

\bibitem[\protect\citeauthoryear{{Luo} \& {Fang}}{{Luo} \&
  {Fang}}{2014}]{luo14}
{Luo} Y.,  {Fang} T.,  2014, \mn@doi [\apj] {10.1088/0004-637X/780/2/170},
  \href {http://adsabs.harvard.edu/abs/2014ApJ...780..170L} {780, 170}

\bibitem[\protect\citeauthoryear{{Nakanishi} \& {Sofue}}{{Nakanishi} \&
  {Sofue}}{2006}]{nak06}
{Nakanishi} H.,  {Sofue} Y.,  2006, \mn@doi [\pasj] {10.1093/pasj/58.5.847},
  \href {http://adsabs.harvard.edu/abs/2006PASJ...58..847N} {58, 847}

\bibitem[\protect\citeauthoryear{{Padoan}, {Cambr{\'e}sy}, {Juvela}, {Kritsuk},
  {Langer}  \& {Norman}}{{Padoan} et~al.}{2006}]{pad06}
{Padoan} P.,  {Cambr{\'e}sy} L.,  {Juvela} M.,  {Kritsuk} A.,  {Langer} W.~D.,
   {Norman} M.~L.,  2006, \mn@doi [\apj] {10.1086/507068}, \href
  {http://adsabs.harvard.edu/abs/2006ApJ...649..807P} {649, 807}

\bibitem[\protect\citeauthoryear{{Pettitt}, {Dobbs}, {Acreman}  \&
  {Bate}}{{Pettitt} et~al.}{2015}]{pet15}
{Pettitt} A.~R.,  {Dobbs} C.~L.,  {Acreman} D.~M.,   {Bate} M.~R.,  2015,
  \mn@doi [\mnras] {10.1093/mnras/stv600}, \href
  {http://adsabs.harvard.edu/abs/2015MNRAS.449.3911P} {449, 3911}

\bibitem[\protect\citeauthoryear{{Pinto}, {Kaastra}, {Costantini}  \&
  {Verbunt}}{{Pinto} et~al.}{2010}]{pin10}
{Pinto} C.,  {Kaastra} J.~S.,  {Costantini} E.,   {Verbunt} F.,  2010, \mn@doi
  [\aap] {10.1051/0004-6361/201014836}, \href
  {http://adsabs.harvard.edu/abs/2010A%26A...521A..79P} {521, A79}

\bibitem[\protect\citeauthoryear{{Pinto}, {Kaastra}, {Costantini}  \& {de
  Vries}}{{Pinto} et~al.}{2013}]{pin13}
{Pinto} C.,  {Kaastra} J.~S.,  {Costantini} E.,   {de Vries} C.,  2013, \mn@doi
  [\aap] {10.1051/0004-6361/201220481}, \href
  {http://adsabs.harvard.edu/abs/2013A%26A...551A..25P} {551, A25}

\bibitem[\protect\citeauthoryear{{Sala} \& {Greiner}}{{Sala} \&
  {Greiner}}{2006}]{sal06}
{Sala} G.,  {Greiner} J.,  2006, The Astronomer's Telegram, \href
  {http://cdsads.u-strasbg.fr/abs/2006ATel..791....1S} {791, 1}

\bibitem[\protect\citeauthoryear{{Wang} \& {Chakrabarty}}{{Wang} \&
  {Chakrabarty}}{2004}]{wan04}
{Wang} Z.,  {Chakrabarty} D.,  2004, \mn@doi [\apjl] {10.1086/426787}, \href
  {http://cdsads.u-strasbg.fr/abs/2004ApJ...616L.139W} {616, L139}

\bibitem[\protect\citeauthoryear{{Wilson}, {Jefferts}  \& {Penzias}}{{Wilson}
  et~al.}{1970}]{wil70}
{Wilson} R.~W.,  {Jefferts} K.~B.,   {Penzias} A.~A.,  1970, \mn@doi [\apjl]
  {10.1086/180567}, \href {http://adsabs.harvard.edu/abs/1970ApJ...161L..43W}
  {161, L43}

\bibitem[\protect\citeauthoryear{{den Herder} et~al.,}{{den Herder}
  et~al.}{2001}]{den01}
{den Herder} J.~W.,  et~al., 2001, \mn@doi [\aap] {10.1051/0004-6361:20000058},
  \href {http://adsabs.harvard.edu/abs/2001A%26A...365L...7D} {365, L7}

\bibitem[\protect\citeauthoryear{{in't Zand}, {Kuulkers}, {Verbunt}, {Heise}
  \& {Cornelisse}}{{in't Zand} et~al.}{2003}]{int03}
{in't Zand} J.~J.~M.,  {Kuulkers} E.,  {Verbunt} F.,  {Heise} J.,
  {Cornelisse} R.,  2003, \mn@doi [\aap] {10.1051/0004-6361:20031586}, \href
  {http://cdsads.u-strasbg.fr/abs/2003A%26A...411L.487I} {411, L487}

\makeatother
\end{thebibliography}

\bsp	
\label{lastpage}
\end{document}